\documentclass[12pt]{article}
\usepackage{amsfonts}
\usepackage{bm}
\usepackage{graphicx}
\usepackage{multicol}

\begin{document}

\title{
Lee-Yang zeros and two-time spin correlation function
}

\maketitle

\centerline {Kh. P. Gnatenko \footnote{E-Mail address: khrystyna.gnatenko@gmail.com}, A. Kargol \footnote{E-Mail address: akargol@hektor.umcs.lublin.pl}, V. M. Tkachuk \footnote{E-Mail address: voltkachuk@gmail.com}}
\medskip
\centerline {\small  $^{1,3}$ \it  Ivan Franko National University of Lviv, }
\centerline {\small \it Department for Theoretical Physics,12 Drahomanov St., Lviv, 79005, Ukraine}

 \centerline {  \small  $^{2}$ \it Instytut Matematyki, Uniwersytet Marii Curie-Sklodowskiej,}
\centerline {\small \it 20-031 Lublin, Poland}

\begin{abstract}
The two-time correlation function for probe spin interacting with spin system (bath) is studied.
We show that zeros of this  function correspond to zeros of partition function of spin system
in complex magnetic field.
The obtained relation gives  new possibility to observe the  Lee-Yang zeros experimentally. Namely, we show that measuring of the time dependence of correlation function allows direct experimental observation of the Lee-Yang zeros.

{\small \sl {\bf Keywords:} Spin system, Lee-Yang zeros, two-time correlation function}\\
PACS numbers: 05.30.-d, 64.60.De
\end{abstract}

\section{Introduction}

After works of Lee, Yang  \cite{Yang52,Lee52} and Fisher  \cite{Fish65} analysis of partition function zeros are considered as a standard tool of studying properties of phase transitions in different systems \cite{Wu}.

It is well known that the partition function of physical system is positive and can not be equal to zero.
 The partition function may have zeros  if we allow the parameters in the hamiltonian of a system to be complex. These zeros are called Lee-Yang zeros.
Lee and Yang studied zeros of partition function for ferromagnetic Ising model with complex magnetic field \cite{Lee52}
and proved the theorem that all zeros are purely imaginary. This theorem
holds for any Ising-like model with ferromagnetic interaction \cite{Lieb81} (see
also \cite{Koz97,Koz03,Koz99}). Later Fisher
generalized the Lee-Yang result to the case of complex temperature \cite{Fish65}.

 Studies of zeros of partition function for spin systems have attracted much attention (see, for instance, \cite{Binek98,Wei12,Wei14,Peng15,Kra15,Kra16} and references therein). At the same time there are essentially smaller number of papers devoted to studies of zeros of partition function of Bose systems (see, for instance, \cite{Mul01,Dij15,Borrmann,Gna17}) and Fermi systems (see, for instance,\cite{Bha11,Zvyagin}).

 It is worth noting that  studies of partition function zeros are important fundamentally. Zeros of partition function fully determine the analytic properties of free energy and are very useful for studies thermodynamical properties of many-body systems.

Because of difficulties in the experimental realization of a many-body system with complex parameters, for a long time studies of Lee-Yang zeros were only theoretical.
In 1998 an experimental access to study the density function of zeros on the Lee-Yang circle for
a ferromagnet was provided \cite{Binek98}. Later in paper \cite{Wei12}(see also \cite{Wei14}),  analyzing decoherence of probe spin, the possibility of direct experimental observation of Lee-Yang zeros for partition function of spin system was shown.
The authors of \cite{Peng15} reported direct experimental observation of Lee-Yang zeros.
In  \cite{Flindt13} it was suggested that dynamical phase transitions
may be also analyzed and detected within the framework of Lee-Yang zeros.
 Report on experimental determination of the dynamical Lee-Yang zeros was presented in \cite{Bran17}.
In our recent paper \cite{Gna17} we showed the possibility of experimental observation of Lee-Yang zeros for interacting Bose gas, considering time-dependent correlation function. In \cite{Gna172} we found that zeros of time-dependent correlation functions of  q-deformed Bose gas are related with the Fisher zeros.

In the present paper
we relate zeros of two-time correlation function of probe spin with zeros
of partition function of spin system (bath) with which probe spin is  interacted.
This relation in principle gives a new possibility
for experimental observation of Lee-Yang zeros for spin systems in addition to that presented in \cite{Wei12,Peng15}.

The paper is organized as follows. In section 2 we give the preliminary information
about the system under consideration. In Section 3 we find relation of zeros of two-time spin-1/2 correlation functions
with Lee-Yang zeros. The possibility of simple experimental realization of considered system is shown in section 4.
Conclusions are presented in Section 5.

\section{Hamiltonian of system under consideration}

We are interested in study of the Lee-Yang zeros of spin-1/2 system under magnetic field with sufficiently general Hamiltonian
\begin{eqnarray}\label{Hamilt}
H=H'-h\sum_{i=1}^N\sigma^z_j,
\end{eqnarray}
and only one restriction that $H'$ commutes with the total spin
\begin{eqnarray}\label{comm}
[H',\sum_{i=1}^N\sigma^z_j]=0.
\end{eqnarray}
Hamiltonian $H^{\prime}$ can be Ising Hamiltonian $H'=-\sum_{jj'}J_{jj'}\sigma^z_j\sigma^z_{j'}$, Heisenberg Hamiltonian
$H'=-\sum_{jj'}J_{jj'}({\bm \sigma}_j{\bm \sigma}_{j'})$ or some other that satisfy (\ref{comm}).
Pauli operators $\sigma^{\alpha}_j$ ($\alpha=x,y,z,$ or $1,2,3$) are related with the spin operators $s^{\alpha}_j=\hbar\sigma^{\alpha}_j/2$ and satisfy the following commutation relations
\begin{eqnarray}
[\sigma^{\alpha}_j,\sigma^{\beta}_{j'}]=2i\delta_{jj'}\epsilon^{\alpha \beta \gamma}\sigma^{\gamma}_j.
\end{eqnarray}
In addition these operators satisfy anticommutation relation
\begin{eqnarray}
\{\sigma^{\alpha}_i, \sigma^{\beta}_i\}=2\delta^{\alpha\beta}.
\end{eqnarray}
Here $\epsilon^{\alpha \beta \gamma}$ is antisymmetric tensor and $\delta_{jj'}$, $\delta^{\alpha\beta}$ are Kronecker
symbols.

We include into consideration probe spin-1/2 coupled to the considered system
(bath) which is described by Hamiltonian (\ref{Hamilt}), with the probe-bath interaction
\begin{eqnarray}
H_{int}=-\lambda\sigma^z_0 \sum_{j=1}^N\sigma^z_j,
\end{eqnarray}
where $\sigma^z_0$ corresponds to the probe spin, $\lambda$ is a coupling constant. So, the total Hamiltonian
is
\begin{eqnarray}\label{TotalH}
H_T=H+H_0+H_{int},
\end{eqnarray}
here $H_0=-h_0\sigma^z_0$ is the Hamiltonian of probe spin placed in magnetic field $h_0$.
Hamiltonian similar to (\ref{TotalH}) was considered in \cite{Wei12}.

\section{Two-time spin correlation function}
Let us consider two-time correlation function for probe spin, which reads
\begin{eqnarray}
\langle \sigma_0^+(t+\tau)\sigma_0^-(t) \rangle={1\over Z_T}{\rm Tr}e^{-\beta H_T}\sigma_0^+(t+\tau)\sigma_0^-(t),
\end{eqnarray}
where, $\sigma_0^{\pm}=(\sigma_0^x\pm i\sigma_0^y)/2$, $Z_T={\rm Tr}e^{-\beta H_T}$ is partition function of the total Hamiltonian, ${\rm Tr}={\rm Tr}_{1,2,...,N} {\rm Tr}_0$ is going over all spins including the probe one.
Heisenberg representation for operators reads
\begin{eqnarray}
\sigma_0^-(t)=e^{iH_T t/\hbar}\sigma_0^-e^{-iH_T t/\hbar}=\nonumber\\
e^{-i\lambda\sigma^z_0 \sum_{j=1}^N\sigma^z_jt/\hbar-ih_0\sigma^z_0 t/\hbar}
\sigma_0^-
e^{i\lambda\sigma^z_0 \sum_{j=1}^N\sigma^z_j t/\hbar+ih_0\sigma^z_0 t/\hbar},
\end{eqnarray}
here we take into account that the probe spin operators commute with other ones.
In addition taking into account that $\sigma_0^-$ anticommutes with $\sigma_0^z$, $\{\sigma_0^-,\sigma_0^z\}=0$, we find
\begin{eqnarray}
\sigma_0^-(t)=\sigma_0^-e^{i2\lambda\sigma^z_0 \sum_{j=1}^N\sigma^z_j t/\hbar+i2h_0\sigma^z_0 t/\hbar}.
\end{eqnarray}
The conjugated operator to $\sigma_0^-(t)$ reads
\begin{eqnarray}
\sigma_0^+(t)=e^{-i2\lambda\sigma^z_0 \sum_{j=1}^N\sigma^z_j t/\hbar-i2h_0\sigma^z_0 t/\hbar}\sigma_0^+,
\end{eqnarray}
The product of these operators can be written as
\begin{eqnarray}
\sigma_0^+(t+\tau)\sigma_0^-(t)=\nonumber\\
=e^{-i2\lambda\sigma^z_0 \sum_{j=1}^N\sigma^z_j (t+\tau)/\hbar-i2h_0\sigma^z_0 (t+\tau)/\hbar}\sigma_0^+\sigma_0^-
e^{i2\lambda\sigma^z_0 \sum_{j=1}^N\sigma^z_j t/\hbar+i2h_0\sigma^z_0 t/\hbar}=\nonumber\\
={1\over 2}(1+\sigma^z_0)e^{-i2\lambda\sigma^z_0 \sum_{j=1}^N\sigma^z_j \tau/\hbar-i2h_0\sigma^z_0 \tau/\hbar}.
\end{eqnarray}

Then correlation function reads
\begin{eqnarray}
\langle \sigma_0^+(t+\tau)\sigma_0^-(t)\rangle= \nonumber \\
={1\over Z_T}{\rm Tr}_{1,2,...,N} {\rm Tr}_0e^{-\beta H_T}{1\over 2}(1+\sigma^z_0)
e^{-i2\lambda\sigma^z_0 \sum_{j=1}^N\sigma^z_j \tau/\hbar-i2h_0\sigma^z_0 \tau/\hbar}=\nonumber\\
={1\over Z_T}{\rm Tr}_{1,2,...,N} e^{-\beta (H'-h\sum_{i=1}^N\sigma^z_j)}{\rm Tr}_0{1\over 2}(1+\sigma^z_0)
e^{(\beta\lambda -i2\lambda \tau/\hbar)\sigma^z_0 \sum_{j=1}^N\sigma^z_j+(\beta h_0-2ih_0\tau/\hbar)\sigma^z_0},
\end{eqnarray}
here we substitute the total Hamiltonian given by (\ref{TotalH}).

Taking  trace over the probe spin and using identity
\begin{eqnarray}
{\rm Tr}_0{1\over 2}(1+\sigma^z_0)e^{\sigma^z_0 \hat A}=e^{\hat A},
\end{eqnarray}
where $\hat A$ is an operator which commutes with $\sigma^z_0$,
we obtain very interesting result
\begin{eqnarray}\label{CorPart}
\langle \sigma_0^+(t+\tau)\sigma_0^-(t)\rangle={e^{(\beta h_0-2ih_0\tau/\hbar)}\over Z_T}{\rm Tr}_{1,2,...,N}
e^{-\beta (H'-\tilde h\sum_{j=1}^N\sigma^z_j)}=\nonumber\\=e^{(\beta h_0-2ih_0\tau/\hbar)}{Z(\beta,\tilde h)\over Z_T},
\end{eqnarray}
here $Z(\beta,\tilde h)$ is the partition function of Hamiltonian (\ref{Hamilt}) with complex magnetic field
\begin{eqnarray} \label{compField}
\tilde h=h+\lambda -i{2\lambda\tau\over \beta\hbar }.
\end{eqnarray}

Thus, we find the direct relation (\ref{CorPart}) of two-time correlation function with the partition function of spin system in complex magnetic field.
According to (\ref{CorPart}) zeros of correlation function are  zeros of partition function $Z(\beta,\tilde h)$ of Hamiltonian (\ref{Hamilt}) with complex magnetic field.

\section{Triangle spin cluster}
In this section we consider one of the possible simple realizations of the total Hamiltonian considered in the previous sections.
Let us study triangle spin cluster the Hamiltonian of which can be associated with the total Hamiltonian $H_T$. Namely, the Hamiltonian
\begin{eqnarray}
H_T=-J\sigma_0^z\sigma_1^z-J\sigma_0^z\sigma_2^z-J\sigma_1^z\sigma_2^z -h(\sigma_0^z+\sigma_1^z+\sigma_2^z)=\nonumber\\
=-J\sigma_1^z\sigma_2^z- h(\sigma_1^z+\sigma_2^z) -h\sigma_0^z-J\sigma_0^z(\sigma_1^z+\sigma_2^z)
\end{eqnarray}
can be rewritten in the form (\ref{TotalH}) with
\begin{eqnarray}\label{Twospin}
H=-J\sigma_1^z\sigma_2^z- h(\sigma_1^z+\sigma_2^z),\\
H_0=-h\sigma_0^z,\\
H_{\rm int}=-J\sigma_0^z(\sigma_1^z+\sigma_2^z),
\end{eqnarray}
where $\lambda=J$ and $h_0=h$. Thus in this case one of the spins (we choose this spin to be $\sigma_0$) can be considered as the probe spin (see Fig. 1).
\begin{figure}[h!]
\includegraphics[width=0.5\textwidth]{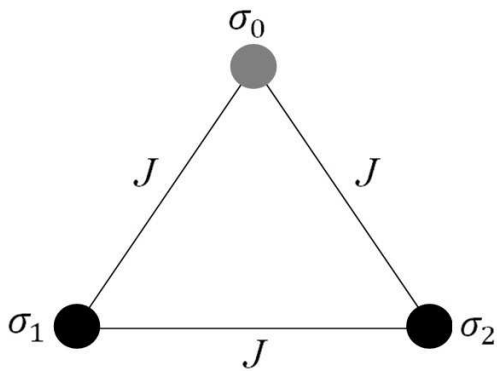}{}
\caption{Triangle spin cluster with $\sigma_0$ being the probe spin. Time correlation function of the probe spin
is proportional to partition function of two spins $\sigma_1$ and $\sigma_2$ in complex magnetic field.}
\label{f1}
\end{figure}
According to (\ref{CorPart}) the time correlation function for spin $\sigma_0$ is related
with partition function of two other spins described by Hamiltonian (\ref{Twospin})
in complex magnetic field (\ref{compField}). The partition function of this Hamiltonian reads
\begin{eqnarray}
Z(\beta,\tilde h)=e^{\beta (J-2 \tilde{h})}\left(q^2+2q e^{-2\beta J}+1\right),
\end{eqnarray}
where we introduce the notation
\begin{eqnarray}
q=e^{2\beta \tilde h}.\label{q1}
\end{eqnarray}
This partition function has zeros $Z(\beta,\tilde h)=0$ at the  points
\begin{eqnarray} \label{q}
q_{\pm}=-e^{-2\beta J}\pm\sqrt{e^{-4\beta J}-1}.
\end{eqnarray}
For antiferromagnetic interaction $J<0$  solutions (\ref{q}) are real,
for ferromagnetic interaction $J>0$ we have the complex solutions
\begin{eqnarray} \label{qc}
q_{\pm}=-e^{-2\beta J}\pm i\sqrt{1-e^{-4\beta J}}.
\end{eqnarray}
Let us consider this case in details. One can verify that $|q|=1$. So, the solutions (\ref{qc}) can be represented in the form
\begin{eqnarray}
q_{\pm}=e^{i\phi}, \ \ \phi=\mp\tan\sqrt{e^{4\beta J}-1} +2\pi n, \ \ n=0,\pm1,\pm2,....\label{rez}
\end{eqnarray}
It means that zeros are achieved at purely imaginary magnetic field that is in agreement with the Lee-Yang theorem.

Taking into account (\ref{compField}), definition of $q$ (\ref{q1}) and (\ref{rez})  we find
that zeros of partition function and zeros of correlation function are achieved at
$h=-J$ and
\begin{eqnarray}\label{timeZero}
\tau={\hbar\over 4J}\left(\pm\tan\sqrt{e^{4\beta J}-1} +2\pi n\right).
\end{eqnarray}
 So, the time correlation function has zeros at the moments of time given by (\ref{timeZero}).

Note that considered in this section simple Ising model describes the behavior of spins for instance in a Molecular Dysprosium Triangle \cite{Luz08}.
So, measuring of time correlation function in this system allows experimental observation of the Lee-Yang zeros.

\section{Conclusions}

The main result of this paper is presented by formula (\ref{CorPart}) which relates the two-time spin-1/2 correlation function with the partition function of spin system in complex magnetic field. The imaginary part of magnetic field according to (\ref{compField}) is related with the time of evolution.
Thus measure of time correlation function allows experimental observation of Lee-Yang zeros.
The Lee-Yang theorem states that zeros of partition function for ferromagnetic spin system
lie  on imaginary axis of magnetic field $\tilde h$. Thus, the zeros of correlation function and respectively the zeros of partition function are achieved at ${\rm Re} \tilde h=0$, that means that $h=-\lambda$. At this condition the correlation function
has zeros at different moments of time that correspond to zeros of partition function. So, measuring of time dependence of the correlation function allows direct experimental observation of Lee-Yang zeros.
This result provides new experimental possibilities in studies of  Lee-Yang zeros in addition to that presented in \cite{Wei12,Peng15}, where zeros were related with decoherence.

We show the possibility of experimental realization of spin systems considered in this paper. One of the simplest possible realization is triangle Ising cluster model which describes the behavior of spins in a Molecular Dysprosium Triangle \cite{Luz08}. A spin of this system can be considered as probe spin. Time correlation function of probe spin in this case is proportional to partition function of two other spins in complex magnetic field.
Zeros of the time correlation function are achieved at the moments of time given by (\ref{timeZero}). Observation of these zeros is equivalent to observation of Lee-Yang zeros.
The realization of this new possibility of experimental observation of Lee-Yang zeros is related with recent progress in experimental measurements of the time
correlation function \cite{Kna13,Ped14,Uhr17,Xin17}, in particular, spin  time correlation functions.
We hope that these methods of measurements of spin correlation function can be applied to the system considered in present paper that allows new possibility for direct experimental observation of Lee-Yang zeros.

\section*{Acknowledgments}
This work was supported in part by the European Commission under the project STREVCOMS PIRSES-2013-612669 and by the State Found for Fundamental Research under the project F76.  The authors thank Prof. Yu. Kozitsky, Prof. Yu. Holovatch and Dr. M. Krasnytska for useful comments and discussions.

\end{document}